\documentclass[]{krakproc}
\usepackage{graphicx}
\begin{document}

%%%%%%%%%%%%%%%%%%%%%%%%%%%%%%%%%%%%%%%%%%%%%%%%%%%%%%%%%%%%%%%%%%%%%%%%%
% version: 01.02.00.
\input{epsf}

\newcommand{\3}{\ss}
\newcommand{\n}{\noindent}
\newcommand{\eps}{\varepsilon}
\def\be{\begin{equation}}
\def\ee{\end{equation}}
\def\ba{\begin{eqnarray}}
\def\ea{\end{eqnarray}}
\def\de{\partial}
\def\div{\nabla\cdot}
\def\grad{\nabla}
\def\rot{\nabla\times}
\def\eg{{\it~e.g.~}}
\def\ie{{\it~i.e.~}}
\def\etal{{\it et~al.~}}
\def\msun{{M_\odot}}
\def\ltsima{$\; \buildrel < \over \sim \;$}
\def\simlt{\lower.5ex\hbox{\ltsima}}
\def\gtsima{$\; \buildrel > \over \sim \;$}
\def\simgt{\lower.5ex\hbox{\gtsima}}

\title{Dusty plasmas on galactic scales}
\author{Yuri A. Shchekinov}
\institute{Department of Physics, University of Rostov,\\
        Rostov on Don, 344090 Russia}
\markboth{Yu. A. Shchekinov}{Dusty plasmas \ldots}

\maketitle

\begin{abstract}
Interstellar dust is spread in galaxies over large scales, often far beyond
the stellar disks. Several mechanisms can be responsible for carrying the
dust both in the vertical and radial directions, producing in general
different spatial dust distributions. In spite of the small mass fraction 
dust can be in some cases dynamically active, and its role in star formation
process and dynamics of shock waves is still not completely understood.
In this review I discuss briefly the issues connected with the large scale
distribution of the interstellar dust, the impact of the dust on star formation 
and on dynamics of shock waves.
\end{abstract}

%%%%%%%%%%%%%%%%%%%%%%%%%%%%%%%%%%%%%%%%%%%%%%%%%%%%%%%%%%%%

%\date{}
%\maketitle

\section{Introduction}

Interstellar dust grains are normally charged with the charge varying
in a wide range from negative to positive, and reaches in diffuse interstellar
environment hundreds of charge units~(Weingartner \& Draine 2001a, Yan et al. 2004).
The characteristic charging time of grains is short, normally of $10^3-10^5$ s,
what means that significant fluctuations of charge from the mean value are rare,
except conditions in molecular clouds, where the mean dust charge is $-1<\langle
z_d\rangle< 0$ and part of time grains spent as neutral. Therefore, 
in a wide range of interstellar parameters dust particles are highly 
charged and involved into overall electrodynamics of the interstellar 
medium (ISM). Charge separation is
important, though, only on short scales of relevant Debye lengths, the
longest being
the dust Debye length $\lambda_d=\sqrt{k_BT_d/4\pi n_dz_de^2}=6\times
10^7T_{e,4}^{1/2}z_{d,2}^{-1/2}
n^{-1/2}$ cm; $T_d=T_e$ is assumed\footnote{$A_m=A/10^m$ in cgs}. Thus,
in most cases quaineutrality condition $n_e=z_in_i+z_dn_d$ is essential 
when the large scale dynamics of a dusty ISM is concerned.
The dust gyrofrequency $\omega_{dc}=10^{-10}z_{d,2}m_{d,10}^{-1}
B_{\mu{\rm G}}$ Hz, where $m_{d,10}=10^{10}m_{\rm H}$, and thus dust
particles are tightly coupled to magnetic lines. Note, that
$z_{d,2}/m_{d,10}$ varies from $\sim 10^{-2}$ for the 
larger grains ($a\simeq 1
\mu$m) to $\sim 10^4$ for the smallest ones ($a\simeq 10$\AA), with the 
corresponding variations of the gyrofrequency 
$(10^{-12}-10^{-6})B_{\rm \mu G}$ Hz. The contribution of
the charged dust into current in the ISM is negligible $j_d/j_e\sim 10^{-10}
|z_{d,2}|$, however when dust is streaming along magnetic lines it
makes electromagnetic waves in the long wavelenght limit 
($\lambda>1$ pc) unstable on relatively short times,
and thus can be a source of MHD turbulence in the ISM~
(Shchekinov \& Kopp, 2004).

On microscopic scales charged dust grains can control chemical structure of
forming protostellar cores through ambipolar diffusion of dust and associated
evacuation of depleted elements outwards~(Ciolek \& Mouschovias, 1996);
it can suppress magnetorotational instability in accretion protostellar disks
and establish a layered accretion with a reduced accretion mass rate of 
refractory elements onto the central star~(Gammie, 1996).
On larger scales charged dust is found to be involved into turbulent MHD
motions, in which they can be accelereted up to $v\sim 10$ km s$^{-1}$,
so that grain-grain collisions can efficiently re-eject refractory elements
into the ISM (Yan et al., 2004). When we turn to the galactic scales
the following aspects of the dynamics of charged dust particles are of
principal significance for the interstellar matter: {\it i)} 
transport of dust particles and associated transport of heavy elements 
frozen on their surface; 
{\it ii)} sedimentation of dust in external gravitation and
radiation fields and possible effects on star formation;
{\it iii)} processing of dust by interstellar shock waves, destruction 
of dust grains and overall circulation of dust in the ISM.

\section{Transport of dust}

%\subsection*{ }
\subsection{Transport in $z$}

It was recognized during the last decade that dust is spread in galaxies
both in vertical and radial directions over much larger scales than it was
thought earlier. In NGC 891 dust is found quite far away from the galactic
plane: $z>0.4$ kpc and in some cases it extends up to $\sim 2$ kpc (Howk \&
Savage 1996, 1999, Rossa et al. 2004), with the total mass of gas in the 
extraplanar
dusty structures of $\sim 2\times 10^8\msun$ assuming local dust-to-gas ratio.
Individual dusty structures are as massive as $\simgt 10^5-10^6\msun$, and
the associated potential energy $\simgt 10^{52}-10^{53}$ erg. Similar dust
distribution was recently reported by Thompson et al. (2004) for NGC 4212.

\subsubsection{SNe}

Multiple SNe explosions are commonly thought to be the principal energy source for
the ejection of gas and dust into galactic haloes through chimney outflows (Norman \&
Ikeuchi, 1989). The required energy for a chimney to work is more than $10^2$ SNe
in normal conditions, i.e. in the Milky Way ISM at solar radius (e.g., Mac Low et al.,
1989, Hensler et al. 1996), although it is sensitive to the gas scale height
$h$ and its velocity dispersion $c_s$: $E\propto c_s^2h^3$, and can vary
substantially. The fraction of OB associations which can produce such powerful
events is small -- for a power law luminosity function of OB associations
as in the Milky Way it is only 0.04 (Shchekinov, 1996, Dettmar et al., 2004).
Moreover, the mass of gas ejected into the halo by a single blowout is
normally $M_e\sim (5-10)\rho_0h^3\simlt 3\times 10^3\msun$, where $\rho_0$ is
the midplane density (Mac Low et al., 1989). This is obviously much smaller than
the mass of dusty clumps observed in NGC 891 and NGC 4212, and therefore these
clumps can be produced by the SNe activity only partly, while the rest is apparently
connected with other transport mechanisms. One should stress in this respect
that dust-to-gas ratio in extraplanar dusty clouds in NGC 891 can be significantly
enhanced, as Dettmar et al. (2004) argue from the non-detection of H$\alpha$ rims
around them. If this is the case, these structures must be formed by
the selective action of the radiation pressure on dust.

\subsubsection{Radiation pressure}

Interstellar radiation field above galactic planes is obviously anisotropic, and
should act on dust particles expelling them in vertical direction
(Barsella et al., 1989, Ferrara et al. 1990, 1991, Shustov \& Vibe, 1995). It
works however only under certain conditions. First, as interstellar dust is
mostly charged it
can move vertically only if the interstellar magnetic field has a significant component
perpendicular to the plane. In principle, the Parker instability can produce
large scale magnetic loops extending vertically up to several kpc
(Kahn \& Brett, 1993, Kamaya et al., 1996, 
Hanasz \& Lesch, 2000, Steinacker \& Shchekinov, 2001), in which case such magnetic loops 
can serve as conduits for the radiation-driven dust. Note, that radiation pressure itself is apparently not capable
to trigger the large scale Parker instability. This is connected with the
fact that the typical value of the volumetric radiation force
$f_{\rm r}\sim \sigma\Phi n_d/c\sim 3\times 10^{-35}$ erg cm$^{-4}$ for
$\Phi\sim 0.01$ erg cm$^{-2}$ s$^{-1}$ is much smaller than the gas or cosmic
ray pressure force $|\grad p|\sim 10^{-33}$ for $|\grad|^{-1}\sim 100$ pc
(Dettmar et al. 2004).

Second condition is connected with a drag force.
The radiation force per dust grain is $F_{\rm r}\sim 3\times 10^{-23}$
dyne with the corresponding acceleration of $a_{\rm g}\sim 3\times 10^{-9}$
cm s$^{-2}$ comparable to that produced by gravitation. 
In the midplane where the gas density is high, $n\simgt 0.1$ cm$^{-3}$, 
the collisional
coupling of the dust and gas is fairly 
strong, and the radiation pressure can produce only a
slow drift of the dust with respect to the gas along ${\bf B}$ -- the drift velocity
is $v_{\rm d}\simlt 0.3-0.5$ km s$^{-1}$ (Weingartner \& Draine, 2001b).
Such a slow motion requires $\sim$Gyr time
scales for the dust to be elevated up to 0.3-1 kpc above the galactic plane.
However, once dust is carried out by a hydrodynamic flow (e.g. convection,
or relatively weak SNe explosions) to distances $z\simgt 0.3$ kpc, collisions
become less frequent and dust particles can be accelerated by radiation 
efficiently.
Depending on the anisotropic radiation flux they reach the 
distance $z\sim 1-3$ kpc
in one to several hundreds Myr, and as their motion
along the loop is converging they form dense clumps in the top parts of
the magnetic loops with overabundant dust (Dettmar et al., 2004). The characteristic
time scale for the radiative dust transport varies as $t_{\rm R}\propto \Phi^{-1/2}$,
and for NGC 891, where the radiation flux can be as high as 
$\sim 0.1$ erg cm$^{-2}$ s$^{-1}$, it is $t_{\rm R}\sim$ 100 Myr, 
while for the conditions in the Milky
Way it is 400 Myr and may be too long compared to the lifetime of Parker
magnetic loops.

\subsubsection{Vortices and convection}

Differential rotation of galactic disks stores kinetic energy of $\sim 10^{57}$ erg,
which can produce the total mechanical luminosity of $\sim
3\times 10^{42}$ erg s$^{-1}$ when being released through hydrodynamical instabilities of sheared flows
in spiral density waves with the growth rate $\gamma\sim |du/dr|\sim
3\times 10^{-15}$ s$^{-1}$. This is factor of 3 higher than the 
energy injected by SNe $\sim 10^{42}$
erg s$^{-1}$. It is known that instabilities of sheared flows in external
gravitational field field can result in formation of vortical motions
(known as tornado cyclones) involving a powerful mass transfer in the 
perpendicular
direction (e.g. Snow, 1982). In a simplified self-similar solution for
a galactic tornado flow Shchekinov (2004) estimated the mass ejection rate by a
single tornado as $\dot M_{\rm ss}\sim 0.001-0.01~\msun$ yr$^{-1}$, and the
total mass ejected into the halo in one lifetime $M_{\rm ej}\sim 10^4-10^5
\msun$, which is comparable to the mass of an individual dusty clump inferred 
by Howk \& Savage (1999) and Thompson et al. (2004) for NGC 891 and NGC 4212 
haloes.

Convective motion associated with spiral density waves is a supplementary
source for elevation of dust along with gas into the halo. 
In numerical simulation
of a 3D hydrodynamics of spiral density waves G\'omez \& Cox (2002) found
efficient bore flows extending up to $z\sim 1.5-2$ kpc with characteristic
velocities $\sim 50-60$ km s$^{-1}$, thus providing the mass circulation
rate of up to 1-3 $\msun$ yr$^{-1}$ for the parameters of the Milky Way ISM.

\subsection{Transport in $R$}

Observations of spiral galaxies in FIR show that dust extends in the radial
direction far beyond the stellar or CO disks. Neininger et al. (1996) reported for
NGC 4565 that the dust 1.2mm continuum emission has a radial scale length twice
as the scale length of the CO emission. Bianchi et al. (2000) found for 7 spiral
galaxies observed by ISO that they have the 200$\mu$m emission distributed in 
the radial direction much
wider than the blue light. The corresponding dust-to-star radial
scale length they inferred is $R_d/R_\ast>1.5$. Moreover, in the case 
of NGC 6946 the best fit
model of the FIR spectrum shows $R_d/R_\ast\simgt 3$. In absolute values
the difference between $R_d$ and $R_\ast$ can be as large as 
$3-10$ kpc, and thus
a powerful mechanism for radial transport of dust is wanted.

\subsubsection{Turbulence and radiation pressure}

It is readily seen that turbulence is not sufficiently fast to provide migration
of dust in the radial direction: with the turbulent diffusivity $D_{\rm t}\simeq
Lv/3\sim 10^{26}$ cm$^2$ s$^{-1}$ what corresponds to the characteristic
length $L\sim 100$ pc, and the velocity dispersion
$v\sim 10$ km s$^{-1}$. The rms displacement $\sqrt{\langle \Delta r^2\rangle}
\sim 3$ kpc can be reached in $\sim 5\times 10^{17}$ s.

In outer regions of the galactic disks, interstellar radiation field is
anisotropic. Weingartner \& Draine (2001c) found the anisotropy of radiation at the 
solar circle ranging from 3\% to 21\% for optical (5500\AA) and UV (1500\AA)
starlight. This forces dust particles to drift along
magnetic lines with
the velocity $v_{\rm d}\sim 0.5$ km s$^{-1}$ (Weingartner \& Draine, 2001b).
In the radial direction the radiation forced drift is suppressed by the magnetic
field by a factor $\omega_B\tau_d\simgt 10^2$, where $\omega_B$ is the dust
gyrofrequency, $\tau_d$ is the drag time for a grain. In principle,
as noted by Cho et al. (2003), in a
relatively weak magnetic field MHD turbulence can provide as high dust
diffusivity as the hydrodynamic turbulence does. However, as mentioned above
even the hydrodynamic turbulence is too slow to carry dust over the scales
$\simgt 3$ kpc in reasonable time.

Shaginyan \& Shchekinov (2004) suggested
a combined scenario with the anisotropic radiation pressure accelerating dust particles in
the radial direction outwards in those small scale regions of a turbulent magnetized
ISM, where the random magnetic field is predominantly radial. In
this scheme dust particles accumulate in the local magnetic valleys 
(with respect to the radial direction), where the
radial component vanishes and the magnetic field becomes predominantly
azimuthal. On next stage accumulated dust diffuses into a neighbour turbulent
eddy through reconnection and the radiation drift continues futher. The
characteristic drift velocity in this scenario depends on the radiaton drif
velocity $v_{\rm d}$ and the reconnection time, and when the latter is not
particularly small it is close to $v_{\rm d}$. The corresponding time for
the dust to diffuse over a 3 kpc radial scale is of the order of 1 Gyr.

\subsubsection{Spiral density waves}

Spiral density waves disturb differential rotation of the gas and impart
to it the radial velocity component, thus providing radial migration of
dust particles by the drag and the spiral gravity force. Vorobyov
\& Shchekinov (2004) have explored this possibility for a two-fluid
(gas and dust) system evolving in an external gravitational potential 
from the stellar
disk and dark matter halo. The dust being injected at the initial time
uniformly inside the corotation radius ($r\leq 6.5$ kpc),
forms a spiralwise distribution
in one rotation period of the galaxy with the arms extending outside the
corotaion (Fig. 1). In approximately 1.5 rotation period dust arms reach radial
distances almost twice the corotation radius ($r\simeq 10$ kpc),
with 5-10\% of the initial
dust mass found in this region. Quite important is that the dust arms are
clearly offset of the gas ones, revealing variations of the dust-to-gas
ratio by factor more than 10-30 between the dust and gas arms.
Moreover, a significant
fraction ($\sim$30\%) of the injected dust migrates most of its time
$\sim 1$ Gyr in the region between the gas arms, and thus can survive
against a hostile influence of strong shocks from SNe inside the gas arms.
\begin{figure}[t!]%[ht]
\epsfxsize=5.5truein
\epsfysize=3.3truein
% \resizebox{\hsize}{!}
\epsfbox{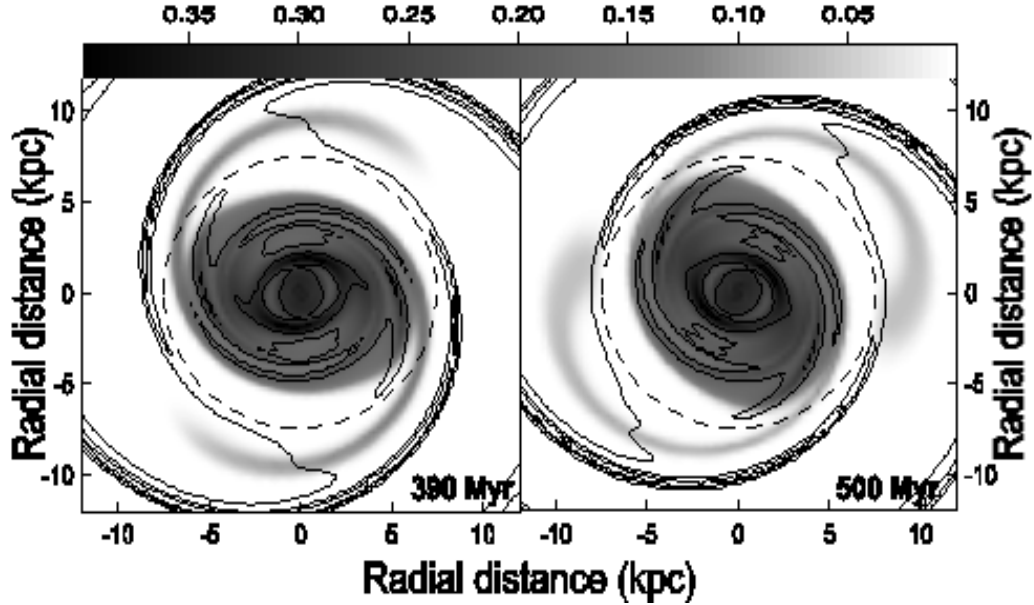}
%   {
%   \includegraphics{figure1.ps}}
 \caption{Contour plot of the gas spiral wave superimposed on the 
gray scale map of the dust spiral wave for 390 and 500 Myr since when the 
dust was injected uniformly inside the corotation radius $r<6.5$ kpc; the dashed 
circle outlines the corotation in the gas disk (from Vorobyov \& Shchekinov, 
2004). The scale bar is in $\msun$ pc$^{-2}$.}
 \label{spirals}
\end{figure}

\section{Star formation}

Dust is known to be an important agent in star formation through its
contribution to thermodynamics and optical characteristics of the gas and
its ability to affect the abundances of the depleted elements in star-forming
regions. At present, it becomes clear that dust can be also dynamically important in
star formation. Theis \& Orlova (2004) have found that a small addition of
a cold dust component (2\% by mass) can strongly destabilize even hot
galactic gaseous disks with a high Toomre parameter $Q=2-3$. The growth rate of
the instability depends on the dust-to-gas ratio, so that nonlinear dust
structures develop in $\sim 13$ galactic rotation periods for dust admixture
of 2\% by mass, and in $\sim 6$ rotation periods for 10\% of dust mass fraction.
The instability results in formation of overdense gas and dust structures
with a charactersitic size in the radial direction of 100-200 pc, which can, 
in principle, give rise to molecular clouds. At present, however, a growing
number of theoretical arguments appeared favouring the interstellar dust to be 
quite a hot component. Yan et al. (2004)
advocate that MHD waves heat dust particles to the velocity dispersion of
10 km s$^{-1}$. This means that in relatively cold HI clouds with
$T\simlt 100$ K, the contribution of such ``hot'' dust particles to pressure
can be comparable with the gas pressure. In these conditions dust can
stabilize gravitational instability and suppress star formation.

An example of physical conditions when dust can be dynamically active in 
triggering star formation was suggested by Cammerer \& Shchekinov (1994). 
In optically thick clouds interstellar UV produces extraheating and an increase 
of pressure in the external layers by photoelectrons from dust grains. This leads 
to squeezing of the cloud and stimulate subsequent gravitational compression. 
This effect was possibly observed by Hester et al. (1996). 

Radiation pressure is known to trigger in a medium with a homogeneous dust
distribution the so-called ``mock gravity'' instability (Field, 1971).
The necessary condition for the instability to grow is that the ratio of 
the radiation to
gas energy density $\rho_{\rm r}/\rho_{\rm g}$
was sufficiently large. In the long-wavelength limit,
$\lambda>L_{\rm d}$ ($L_{\rm d}$ being the extinction length), the required
condition is $\rho_{\rm r}/\rho_{\rm g}>1$, and thus is close to be fulfilled in
the regions with the gas pressure $P\simlt 3\times 10^3$ K cm$^{-3}$.
The characteristic masses for this instability
$M_{\rm \bar g}\sim 10^6\zeta_d^{-3}n^{-2}\msun$, where $\zeta_{\rm d}$ is the
dust-to-gas ratio in units of the mean Milky Way value,
can be reasonable for typical ISM conditions. The growth time, though, is
normally longer than the gravitational free-fall time $t_{\rm ff}$:
$t_{\rm \bar g}\sim 10^8\zeta^{-1}n^{-1}$ yr. In those regions, however, where
dust is overabundant due to the action of selective forces (such as radiation
pressure) $t_{\rm \bar g}$ can decrease considerably and become even shorter
than $t_{\rm ff}$. From this point of view the ``mock gravity'' seems capable 
to act in the extraplanar dusty clouds, such as observed in NGC 891 and
NGC 4212, if they have an enhanced dust-to-gas ratio as advocated by Dettmar
et al. (2004). Note in this connection detection of
several extraplanar regions of recent star formation in the edge-on galaxy
NGC 55 by T\"ullmann et al. (2003). Observations of possible pecularities in
abundances of refractory elements in these HII regions would indicate whether
the radiation pressure contributed to initiation of star formation through the 
selective action on dust particles.

\section{Dust in magnetized shocks}

Interstellar shock waves are the principal source of dust destruction (see
review by Draine 2003). In most cases interstellar dust behind shock waves 
is treated as a passive component having minor effects on dynamics of the 
shocked gas. However, dust grains can be dynamically
active in oblique shocks and can affect their structure and evolution 
through the instability similar to the mirror
instability in plasma. In radiative shocks gas density increases
$\rho\propto T^{-1}$ and results in an encrease of the frozen-in magnetic
field $B_{||}\propto \rho$. Due to the betatron acceleration the transverse
velocity dispersion of dust grows as $v_{d,\perp}^2\propto B_{||}$,
so that the dust component behind the shock can become highly anisotropic
with the transverse dust pressure comparable to the gas one (McKee et al., 1987).
In these conditions oblique perturbations with
$k_\perp\neq 0$ are unstable with the growth rate $\omega\sim
\sqrt{\beta_{d,\perp}}k_{||}v_{d,\perp}$, $\beta_{d,\perp}$ being the partial
transverse dust pressure (Shaginyan \& Shchekinov, 2004).

Another aspect connected with interaction of the shock waves and the interstellar dust 
relates to segregation of dust and variations of the dust-to-gas ratio in 
the ISM. Oblique shock waves can produce spatial separation of the dust and gas. 
When passing through the shock front with a gradiant of magnetic field
parallel to the front, $dB_{||}/dz\neq 0$, dust particles decellerate
and experience mirroring when move along the front. At the stagnation point, 
where their longitudinal kinetic energy vanishes, they accumulate and form dusty troughs
with an enhanced dust-to-gas ratio (Shaginyan \& Shchekinov, 2004).

\section{Conclusions}

A common feature of many mechanisms carrying dust in the ISM 
is that they act selectively,
and result in segregation of dust:
\begin{itemize}

\item{} Interstellar radiation field can be an efficient agent for the dust
transport both in the vertical and radial directions, and as it acts on
the dust particles selectively it produces regions with overabundant dust. 
The origin of the dusty structures in the haloes of NGC 891 and NGC 4212 
can be connected with the radiation pressure.

\item{} Spiral density waves can drive dust in the radial direction outward. 
In this case a combined
action of the gravitational field of the spiral wave and the collisional drag
force result in spatial separation of the dust and gas, and provide
large scale variations of the dust-to-gas ratio.

\item{} Passing through oblique magnetized shocks dust particles can be 
segregated on the scales of nonuniformity of the magnetic field
behind the shock front.
\end{itemize}

\noindent
Interstellar dust can be a dynamically active component in structuring
the ISM:

\begin{itemize}
\item{} Interstellar dust can be dynamically important in regulation of 
star formation process. Depending on the kinetic temperature of the dust 
it can either stimulate gravitational instability when it is cold, 
or can be a suppressing factor if it is kinetically hot. 

\item{} Dust particles heated by the betatron mechanism can destabilize 
interstellar oblique shock waves.

\end{itemize}

\section*{Acknowledgements}

This work was supported by the German Science Foundation (DFG)
within Sonderforschungsbereich 591 TP A6. I thank the organizers for 
the invitation and hospitality.

\end{document}